\documentclass{article}

\usepackage[utf8]{inputenc}
\usepackage{amsmath,amssymb,amsfonts,amsthm,tensor}
\usepackage{mathrsfs}
\usepackage{bbold}
\usepackage{cite}
\usepackage{verbatim}
\usepackage{float}
\usepackage{color}
\usepackage[a4paper, total={6in, 8in}]{geometry}
\usepackage{appendix}
\usepackage{hyperref}
\usepackage{authblk}
\usepackage{upgreek}
\usepackage{stmaryrd}

\newcommand{\snorm}{k_\zeta } 
\newcommand{\knorm}{{k_0 }} 
\newcommand{\amf}{{\omega }}
\newcommand{\ff}[1]{{\left\lfloor{#1}\right\rfloor}}
\newcommand{\cf}[1]{{\left\lceil{#1}\right\rceil}}

\newcommand{\p}[1]{{\left({#1}\right)}}
\newcommand{\comm}[1]{{\left[{#1}\right]}}
\newcommand{\acomm}[1]{{\left\{{#1}\right\}}}

\newcommand{\nn}{\nonumber}
\newcommand{\spindle}{\mathbb{\Sigma}}
\newcommand{\blens}{\mathbb{\Lambda}}
\newcommand{\ads}{{\mathrm{AdS}}}
\newcommand{\orb}{{\mathscr O}}

\newcommand{\fs}{{\rm f}}
\newcommand{\rem}[2]{{\left\llbracket{#1}\right\rrbracket_{#2}}}

\newcommand{\mN}{\mathfrak m_+}
\newcommand{\mS}{\mathfrak m_-}
\newcommand{\nN}{n_+}
\newcommand{\nS}{n_-}
\newcommand{\pN}{\mathfrak p_+}
\newcommand{\pS}{\mathfrak p_-}
\newcommand{\qN}{q_+}
\newcommand{\qS}{q_-}
\newcommand{\tN}{t_+}
\newcommand{\tS}{t_-}
\newcommand{\wN}{w_+}
\newcommand{\wS}{w_-}
\newcommand{\zN}{z_+}
\newcommand{\zS}{z_-}

\newcommand{\dd}{{\rm d}}
\newcommand{\ee}{{\rm e}}

\newcommand{\im}{{\rm i}}

\newcommand{\st}{{ \upsigma}}

\numberwithin{equation}{section}

\title{Orbifold Indices in Four Dimensions}
\author{Antonio  Pittelli}
\affil{\emph{Dipartimento di Matematica, Universit\`a di Torino}, 
\\ \emph{Via Carlo Alberto 10, 10123 Torino, Italy}
\\ \emph{INFN, Sezione di Torino, Via Pietro Giuria 1, 10125 Torino, Italy}}

\begin{document}

\maketitle

\begin{abstract}
\noindent   
We introduce  supersymmetric   indices for    four-dimensional gauge theories defined  on $\mathscr O \times S^1$,  where $\mathscr O $ is a   circle bundle  over  the weighted complex projective line informally known as spindle. Trivial fibrations yield  a four-dimensional  version of  the spindle index, which we obtain by applying    localization   to partition functions of theories   on the direct product of a spindle and a two-dimensional torus. Conversely, non-trivial fibrations lead to the  branched lens index, which we compute by localizing theories on the direct product of a circle and a branched covering of the lens space, possibly  endowed with conical singularities. The branched lens index encompasses the maximally refined four-dimensional lens index as a special case.

\end{abstract}

\newpage

\tableofcontents

\section{Introduction and Summary}

Supersymmetric indices have found numerous compelling applications in both physics and mathematics, owing to their versatile and diverse equivalent formulations. For example, the index of a supersymmetric quantum system evolving over time on a manifold $\mathscr{M}$ can be represented as a flavored Witten index, which is obtained by taking the trace computed over the corresponding Hilbert space $\mathscr{H}\comm{\mathscr{M}}$ and  weighting it by fugacities that encode the quantum numbers of the theory, along with $(-1)^F$, where $F$ is the fermionic-number operator. This definition  portrays the index as a tool for enumerating BPS states and categorizing them based on their representation within the symmetry algebra of the model. Consequently, it provides   group-theoretical insights and deepens  our understanding of the model's non-perturbative behavior, such as dynamical supersymmetry breaking  \cite{Witten:1982df} and hidden algebraic structures \cite{Beem:2013sza}.

Moreover, the same object coincides with the   index of a suitable transversally elliptic differential operator on $\mathscr{M}\times S^1$, which can be computed by means of equivariant integrals or  fixed-point formulae \cite{Drukker:2012sr,Pestun:2016qko,Panerai:2020boq}, methods that have recently been applied also to  supergravity theories \cite{BenettiGenolini:2023kxp,BenettiGenolini:2023yfe,BenettiGenolini:2023ndb,Martelli:2023oqk,Colombo:2023fhu}. This   highlights the   connection between supersymmetric indices,   topological invariants and  characteristic classes arising from the underlying geometry. Eventually, the supersymmetric index of a system quantized on $\mathscr{M}$ can be   expressed as the Euclidean path integral of the theory defined on $\mathscr{M} \times S^1$. This representation facilitates the computation via  supersymmetric localization \cite{Pestun:2007rz}, an  approach that has been successful  across various scenarios, encompassing indices quantized on spheres \cite{Imamura:2011su,Kapustin:2011jm,Closset:2013sxa,Benini:2015noa}, Riemann surfaces \cite{Benini:2016hjo}, manifolds with boundaries \cite{Yoshida:2014ssa,Longhi:2019hdh,Crew:2023tky} and beyond \cite{Nishioka:2014zpa,David:2016onq,Pittelli:2018rpl,Iannotti:2023jji}. These computations were crucial  for validating gauge-gravity dualities \cite{Benini:2012cz,Benini:2013cda,Benini:2015eyy,Benini:2016rke,Cabo-Bizet:2018ehj,Benini:2018mlo,Benini:2018ywd,Benini:2020gjh}. 

Generalizing the aforementioned  achievements, this paper presents  orbifold indices for supersymmetric  gauge theories formulated on $\orb\times S^1$, with  $\mathscr O$ denoting  a three-dimensional orbifold potentially endowed with  conical singularities. This extends to four dimensions the  partition functions of three-dimensional theories on $\spindle\times S^1$ and $\blens$ considered  for the first time in \cite{Inglese:2023wky,Inglese:2023tyc}. Singular orbifolds  are particularly important in the context of the AdS/CFT correspondence, where they play the role of non-trivial saddle points of the gravitational path integral \cite{Ferrero:2020laf,Ferrero:2020twa,Ferrero:2021wvk,Cassani:2021dwa,Ferrero:2021etw,Faedo:2021nub,Ferrero:2021ovq,Cheung:2022ilc,Suh:2022pkg,Boido:2022iye,Boido:2022mbe,Faedo:2022rqx,Amariti:2023mpg,Hristov:2023rel,Amariti:2023gcx,Faedo:2024upq,Boisvert:2024jrl,Ferrero:2024vmz,Macpherson:2024frt}. Our focus lies on maximally refined orbifold indices, thus we investigate cases where $\orb$ are circle bundles over a spindle $\spindle=\mathbb{WCP}^1_{\comm{\nN, \nS}}$, which preserve a $U(1)^3$ isometry on $\orb\times S^1$. Previous studies   explored the case of unrefined indices on circle fibrations over an orbifold Riemann surface \cite{Closset:2018ghr}. If $\orb$ is a trivial circle fibration over $\spindle$,   the topology of $\orb$ is $\spindle\times S^1$ and     Euclidean path integrals of $\mathcal N=1$ gauge theories on topologically twisted $\spindle \times T^2$ represent the four-dimensional uplift of the spindle index introduced in \cite{Inglese:2023wky,Inglese:2023tyc}.  On   twisted $\spindle \times T^2$  a chiral multiplet of R-charge $r$  in a  representation  $\mathfrak R_G$ of a  gauge or flavour group $G$ contributes  to the orbifold index as
\begin{align}\label{eq: 1loopcmspindlet2} 
    Z^{\rm CM}_{\spindle\times T^2} 
    &   = \prod_{\rho \in \mathfrak R_G} e^{2 \pi \im \Psi^{\rm CM}\p{\rho\p{\mathfrak m} , \rho\p{\gamma_G} } } \frac{  \Gamma_e\p{ z^{-\rho} q^{\frac{1}{2}\comm{1 - \mathfrak b\p{\rho\p{\mathfrak m} } }} ; q , p   } }{ \Gamma_e\p{ z^{-\rho}  q^{ \frac{1}{2}\comm{1 + \mathfrak b\p{\rho\p{\mathfrak m} } } } ; q , p }}    ~ ,
\end{align}
  where the integers $\mathfrak m$  parametrize gauge or flavour fluxes in the co-root lattice $\Gamma_{\mathfrak g}$ of the Lie algebra $\mathfrak g$ of $G$,   $z={e^{2 \pi \im \gamma_G}}$ are fugacities for the gauge\footnote{In the partition function, fugacities corresponding to gauge symmetries are integrated over a complex contour $\mathscr C$ fixed e.g. by the Jeffrey-Kirwan procedure \cite{jeffrey1994localization}.} or flavour symmetry $G$,  the phase factor $\Psi^{\rm CM}$ is reported in the main text,   $q=e^{2 \pi \im \amf}$ is the fugacity for the angular momentum on the spindle, $p=e^{2 \pi \im \tau}$ the fugacity for translations on the torus with modular parameter $\tau$, $\Gamma_e$ is the elliptic Gamma function and $\mathfrak b$ is the effective flux introduced in \cite{Inglese:2023wky,Inglese:2023tyc}, which is  related to the degree of the line bundle\footnote{The product $r\p{\nN+\nS}$ need be even for {\rm L} to be well defined.} ${\rm L}=\mathcal O\p{ -  \mathfrak m  - \p{r/2} \p{\nN+\nS}}$ on $\spindle$ via $\mathfrak b = 1 + {\rm deg}\p{\rm L}  $. Subsequently, $Z^{\rm VM}_{\spindle\times T^2} $ is obtained from $Z^{\rm CM}_{\spindle\times T^2} $ after setting $r=2$ and $\mathfrak R_G={\rm adj}_G$.   To ensure consistency, the result (\ref{eq: 1loopcmspindlet2}) was computed by using two distinct techniques, namely the equivariant index theorem on orbifolds \cite{vergne,MEINRENKEN1998240} and the eigenvalue pairing method.  Both approaches yielded identical results, confirming  the reliability  of the  analysis. The representation of partition functions on $\spindle\times T^2$ in terms of a Witten index is 
\begin{align}
	I_{\spindle\times T^2} = {\rm Tr}_{\mathscr H\comm{\spindle\times S^1}}\comm{\p{-1}^{\mathcal F} e^{- 2\pi   \mathcal H} }  = {\rm Tr}_{\mathscr H\comm{\spindle\times S^1}}\comm{\p{-1}^{\mathcal F}  z_i^{-  \mathcal Q_i }   q^{ \mathcal J} p^{-\mathcal P} } ~ ,
\end{align}
 where $\mathscr H\comm{\spindle\times S^1}$ is the Hilbert space of the system on $\spindle\times S^1$, $\mathcal H$ is the Hamiltonian,  $z_i = e^{- 2\pi \im \upphi_i}$ encode gauge, flavour and R-symmetry fugacities; $\mathcal J$ is the generator of angular momentum on the spindle while $\mathcal P$ generates translations on the torus. In particular, the effective R-symmetry fugacity emerging from the localization procedure is $\gamma_R$, satisfying the constraint
 \begin{align}\label{eq: rsymconstspindlet2}
 	& \gamma_R - \amf \frac{\chi_-}{4} - \tau \frac{n_1}{2} =  \frac{n_2}{2} ~ , & \chi_\st = \frac{1}{\nS} + \st  \frac{1}{\nN}  ~ , \qquad  n_1, n_2 \in \mathbb Z ~ ,  
 	\end{align}
 	where $\st=\pm1$ for twist and anti-twist respectively and the integers $n_1, n_2$ are related to the periodicity of spinors on the torus. The calculation of path integrals on  topologically twisted $\spindle\times T^2$ naturally generalizes the results obtained in \cite{Closset:2013sxa} and \cite{Benini:2015noa}, where the partition functions of quantum field theories on topologically twisted $S^2\times T^2$ were computed. These partition functions are intimately linked to the elliptic genus of $\mathcal N=(0,2)$ gauge theories on $T^2$, as discussed in \cite{Witten:1986bf,Benini:2013nda}. We anticipate similar connections between two-dimensional theories on the torus and four-dimensional theories on $\spindle\times T^2$. Holographically, the partition functions of gauge theories on topologically twisted $\spindle\times T^2$ should correspond to supergravity theories with a near-horizon geometry described by $\ads_3 \times \spindle$ with R-symmetry twist, as discussed in \cite{Ferrero:2021etw}.


On the other hand, if $\orb$ is a non-trivial circle bundle over $\spindle$, it generally is  a branched covering of the lens space, which we denote as $\blens$. The orbifold $\blens$ is characterized  by three positive integers $\p{\nN, \nS, n}$, where the first two encode the conical singularities of the spindle base, while the third labels  the flux of the fibration. In particular, the branched lens space becomes a  three-sphere if $n=1$, whereas  $\blens$ reduces to the smooth lens space $L\p{n,1}\equiv S^3/\mathbb Z_n$ if $\nN=\nS = 1$. Partition functions of $\mathcal N=2$ gauge theories on $\blens$ were  calculated   in \cite{Inglese:2023tyc}, while in this article we undertake the computation of the Euclidean path integral of $\mathcal N=1$ theories on $\blens\times S^1$. In this case, the one-loop determinant of a  chiral multiplet transforming  in  $\mathfrak R_G$  is given by the following  ratio of $q$-Pochhammer symbols:
\begin{align}\label{eq: 1loopcmblenss1} 
	Z_{\blens\times S^1}^{\rm CM}  & = \prod_{\rho\in\mathfrak R_G} e^{2 \pi \im \Psi^{\rm CM} \comm{\rho\p{\gamma_G} , \rho\p{\mathfrak h}} } \prod_{j=0}^{n-1}  \frac{ \prod_{i \in \mathbb J_+\p{j, \rho\p{\mathfrak h} } } \p{ q_1^{  \p{1+j}/n  } q_2^{ \p{1+i}/n  }    z^{-\rho} ; q_1, q_2  }_\infty }{ \prod_{k \in \mathbb J_-\p{j, \rho\p{ \mathfrak h} }} \p{ q_1^{  k/n} q_2^{ j/n} z^\rho  ; q_1 , q_2 }_\infty }    ~ ,
\end{align}
where $\mathfrak h=0, \dots, \p{n-1}$  is an integer in $\mathbb Z_n$,     $q_{1,2}=e^{2 \pi \im \amf_{1,2}}$ are fugacities  for the angular momentum on $\blens$ and $\mathbb J_\pm$ are subsets of $\mathbb Z_{n_\pm}$, respectively. The fundamental  degrees of freedom of gauge theories on $\blens\times S^1$ are counted by the Witten index
\begin{align}
	I_{\blens\times S^1} = {\rm Tr}_{\mathscr H\comm{\blens}}\comm{\p{-1}^{\mathcal F} e^{- 2\pi   \mathcal H} }  = {\rm Tr}_{\mathscr H\comm{\blens}}\comm{\p{-1}^{\mathcal F} z_i^{- \mathcal Q_i } q_1^{- \frac{\nN}{n}\mathcal J_\varphi +\tN \mathcal J_\psi } q_2^{- \frac{\nS}{n}\mathcal J_\varphi +\tS \mathcal J_\psi } } ~ ,
\end{align}
with $\mathscr H\comm{\blens}$ being  the Hilbert space of states quantized  on $\blens$,   $\mathcal J_\varphi$ generates rotation of the base of the fibration  $\spindle$  and $\mathcal J_\psi$ represents translations along the $S^1$ fiber. In the case of $\blens\times S^1$ the   effective R-symmetry fugacity fulfils  
\begin{align}\label{eq: rsymconstblenss1}
	& \gamma_R - \frac{\amf_1 + \amf_2}{2 n} = \frac{n_3}{2} ~ , & n_3 \in \mathbb Z ~ ,
\end{align}
where  the integer $n_3$ is even (odd)  if Killing spinors are periodic (anti-periodic)   on the circle.


Promising directions for future investigation encompass exploring the large-$N$ limit of orbifold indices of  $\mathcal N=4$ Yang-Mills theory with $SU(N)$ gauge group, as well as quiver gauge theories with holographic duals. In  the context of $\spindle\times T^2$, such a limit  is expected to establish connections with BPS black strings in $\ads_5$, extending previous findings pertaining to $S^2 \times T^2$ \cite{Closset:2013sxa,Hosseini:2016cyf,Hosseini:2019lkt,Hosseini:2020vgl}. Similarly, exploring  the large-$N$ limit of maximally supersymmetric Yang-Mills on $\blens\times S^1$ should reproduce the entropy of asymptotically $\ads_5$ black holes featuring an $\ads_2 \times \blens$ near-horizon geometry.

 This generalization to orbifolds builds upon prior investigations concerning the four-dimensional superconformal index \cite{Benini:2018mlo,Benini:2018ywd,Cabo-Bizet:2018ehj,Benini:2020gjh,Colombo:2021kbb}, with the constraints (\ref{eq: rsymconstspindlet2}) and (\ref{eq: rsymconstblenss1}) potentially holding interpretations in terms of angular momenta and electromagnetic potential of  the supergravity dual.

Another intriguing avenue involves the first-principles derivation of partition functions of $\mathcal N=1$ gauge theories on $\spindle\times T^2$ with R-symmetry anti-twist on $\spindle$, corresponding to the four-dimensional uplift of the anti-twisted spindle index in three dimensions.  Building on   \cite{Inglese:2023wky}, it may be conjectured that $Z^{\rm CM}_{\spindle\times T^2} |_{\text{anti-twist}} $ is given by the general expression
\begin{align}\label{eq: 1loopcmspindlet2anytwist} 
         Z^{{\rm CM}, \p{\st}}_{\spindle\times T^2}  = \prod_{\rho \in \mathfrak R_G} e^{2 \pi \im \Psi^{{\rm CM}, \p{\st}}\p{\rho\p{\mathfrak m} , \rho\p{\gamma_G} } }   \frac{  \Gamma_e\p{ z^{-\st \rho}  q^{ \frac{\st}{2}\comm{1 - \mathfrak b\p{\rho\p{\mathfrak m} }} } ; q , p }  } { \Gamma_e\p{ z^{-\rho} q^{\frac{1}{2}\comm{ 1 + \mathfrak b\p{\rho\p{\mathfrak m} }  }} ; q , p   }  }  ~ , 
\end{align}
upon setting $\st=-1$. The   constraint fulfilled by the  effective R-symmetry fugacity $\gamma_R$ for an $\mathcal N=1$ supersymmetric theory on $\spindle \times T^2$ with either twist or anti-twist should be  
 \begin{align}\label{eq: rsymconstspindlet2anyt}
 	& \gamma_R - \amf \frac{\chi_{-\st} }{4} - \tau \frac{n_1}{2} =  \frac{n_2}{2} ~ , &  n_1, n_2 \in \mathbb Z ~ ,  
 	\end{align}
with $\st=\pm1$ corresponding to twisted and anti-twisted $\spindle \times T^2$, respectively. According to the AdS/CFT duality, the large-$N$ limit  of (\ref{eq: rsymconstspindlet2anyt}) should relate to the supergravity solutions found and investigated in \cite{Ferrero:2020laf,Ferrero:2021etw,Hosseini:2021fge,Boido:2021szx}. However, we could not find solutions to the $\mathcal N=1$ Conformal Killing spinor equations on anti-twisted $\spindle\times T^2$, in agreement with \cite{Arav:2022lzo}. Although the gravitational duals of the direct product between a spindle and a torus have been extensively studied \cite{Ferrero:2020laf,Ferrero:2021etw,Hosseini:2021fge}, a direct validation of (\ref{eq: 1loopcmspindlet2anytwist}) with $\st=-1$ through a localization computation starting from a $\mathcal N=1$ rigid supersymmetric background on $\spindle\times T^2$ with R-symmetry anti-twist on $\spindle$, remains  pending.

Eventually, the computation of the 1-loop determinant on $\spindle\times T^2$ via index theorem inherently decomposes the result into distinct blocks, as delineated in (\ref{eq: ZblockN}) and (\ref{eq: ZblockS}). A similar phenomenon was observed in \cite{Inglese:2023wky,Inglese:2023tyc}. A compelling avenue for further investigation involves probing the relationship between such blocks and partition functions on orbifolds with boundaries, such as $\p{D^2/\mathbb Z_n}\times \p{S^1}^{\times m}$ with $n, m\in \mathbb N^*$ and $D^2$ being a disk or a hemisphere. This would   provide  the orbifold extension of holomorphic blocks \cite{Beem:2012mb}.


\paragraph{Outline.} 


In Section \ref{sec: susybackground} we set up  the background geometry by introducing a convenient line element on $\spindle \times T^2$ and $\blens \times S^1$. Subsequently, we proceed to solve the $\mathcal N=1$ conformal Killing spinor equations on both geometries, finding that  non-trivial $U(1)$ R-symmetry connections are necessary in order to preserve supersymmetry. Furthermore, we determine the background fields necessary for these conformal Killing spinors to satisfy the Killing spinor equations derived from the rigid limit of new minimal supergravity. In Section \ref{sec: susylocalization} we perform the computation of  partition functions for $\mathcal N=1$ gauge theories, consisting of vector and chiral multiplets on $\spindle \times T^2$ and $\blens \times S^1$, by employing supersymmetric localization. Especially, we evaluate one-loop determinants on topologically twisted $\spindle \times T^2$ using two distinct methodologies: the method of eigenvalue pairing and the index theorem on orbifolds. We  corroborate  our conjecture for one-loop determinants on anti-twisted $\spindle \times T^2$ by means of index formulae.


\paragraph{Acknowledgements.} 

We thank Fabrizio Nieri, Dario Martelli and Alberto Zaffaroni for   illuminating discussions and useful comments on a draft of this paper.


 \section{Background Geometry}\label{sec: susybackground}

\subsection{Supersymmetry on $\spindle\times T^2$}

We now derive the $\mathcal N=1$    supersymmetric background on topologically twisted $\spindle\times T^2$. In analogy with \cite{Closset:2013sxa} we employ  
\begin{align}\label{eq: lespindlet2}
    \dd s^2 = L^2 \acomm{ f^2 \dd \theta^2 + \sin^2\theta\p{ \dd \varphi + \Omega_3 \dd x + \Omega_4 \dd y }^2 + \beta^2 \comm{\p{\dd x + \tau_1 \dd y}^2 + \tau_2^2 \dd y^2  } } ~ ,
\end{align}
as a line element  on $\spindle \times T^2$, where $\theta\in\comm{0, \pi}$ and $\varphi\in[0, 2 \pi)$ represent the latitude  and the longitude  on the spindle  $\spindle$, respectively; while  $x,y\in [0, 2 \pi)$ are coordinates on the two circles of the two-torus $T^2$ with modular parameter $\tau = \tau_1 + \im \tau_2 \in \mathbb C$. The constant   $\beta\in\mathbb R$ encodes the  dimension of    $T^2$ in units of $L$, which is the equatorial radius of the spindle. The real parameters  $\Omega_{3,4}$ will appear in the partition function in the form of the complex linear combination $\amf  = \tau \Omega_3 - \Omega_4 $, with $ q  =e^{2 \pi \im \amf}$ playing the role of fugacity for the angular momentum on $\spindle$.  The function $f=f(\theta)$ in (\ref{eq: lespindlet2}) fulfils $ f(0) = \nN $ and  $ f(\pi) = \nS$. Such behaviour is consistent with the tangent space of $\spindle$ being $\mathbb C/n_\pm$ at the north and south pole of the spindle. Furthermore,
\begin{align}
     \frac{1}{4\pi}\int_\spindle \sqrt{g_\spindle} R_\spindle  = \frac{1}{\nS}+ \frac{1}{\nN} = \chi_+ = \chi  ~ ,
\end{align}
 where $\chi$ is the Euler characteristic of the spindle $\spindle=\mathbb{WCP}^1_{\comm{\nN, \nS}}$. In the frame
\begin{align}
    \ee^1 & = L f ~ , \nn\\
    \ee^2 & = L \sin\theta  \p{ \dd \varphi + \Omega_3 \dd x + \Omega_4 \dd y   } ~ , \nn\\
     \ee^3 & = L \beta   \p{ \dd x + \tau_1 \dd y  } ~ , \nn\\ 
     \ee^4 & = L \beta  \tau_2 \dd y   ~ ,
\end{align}
the spinors
\begin{align}
     & \zeta_\alpha   = \snorm e^{\frac{\im}{2}\p{\alpha_2  \varphi  + \alpha_3 x  + \alpha_4  y }} \begin{pmatrix} 1 \\ 0 \end{pmatrix}_\alpha    ~ , & \widetilde \zeta^{\dot\alpha}   =  \snorm e^{-\frac{\im}{2}\p{\alpha_2  \varphi  + \alpha_3 x  + \alpha_4  y }} \begin{pmatrix} 0 \\ 1 \end{pmatrix}^{\dot\alpha}    ~ ,
\end{align}
with $\snorm$ being a normalization and $ \alpha_{2,3,4}$  constant phases, solve  the Killing,
\begin{align}
	& \p{\nabla_\mu - \im A_\mu }\zeta + \im V_\mu \zeta + \im V^\nu \sigma_{\mu \nu}\zeta = 0 ~ ,  & \p{\nabla_\mu + \im A_\mu }\widetilde\zeta - \im V_\mu \widetilde \zeta - \im V^\nu \widetilde  \sigma_{\mu \nu} \widetilde \zeta = 0 ~ ,
\end{align}
 and the Conformal Killing spinor equations
 \begin{align}
	& \p{\nabla_\mu - \im A^C_\mu }\zeta + \frac14 \sigma_\mu \widetilde \sigma^\nu  \p{\nabla_\nu - \im A^C_\nu }\zeta   = 0 ~ ,  & \p{\nabla_\mu + \im A^C_\mu }\widetilde\zeta  + \frac14 \widetilde \sigma_\mu  \sigma^\nu  \p{\nabla_\nu + \im A^C_\nu }\widetilde\zeta   = 0 ~ ,
\end{align}  
 if the background fields
\begin{align}
    A^C & = \frac{1}{2}\p{ \omega_{12} + \alpha_2 \dd \varphi  + \alpha_3 \dd x  + \alpha_4 \dd y   }   ~ , \nn\\
    V  &=  L \beta \kappa \p{ \dd x + \tau \dd y }   ~ , \nn\\
    A  &=  A^C + \frac{3}{2}V  ~ ,
\end{align}
are turned on, where 
\begin{align}
    \omega_{12} & = - f^{-1} \cos\theta \p{ \dd\varphi + \Omega_3 \dd x + \Omega_4 \dd y }  ~ ,
\end{align}
is a non-trivial component of the spin-connection.   The flux of the R-symmetry connection satisfies
\begin{align}
    \mathfrak f_R & = \frac{1}{2 \pi }\int_\spindle \dd  A = \frac{1}{2}\p{  \frac{1}{\nS} + \frac{1}{\nN} } = \frac{\chi_+}{2}  = \frac{\chi}{2}   ~ ,
\end{align}
as it behooves the topological twist on $\spindle$. If $\alpha_{3,4}\in \mathbb Z$, then $\zeta$ and $\widetilde \zeta$ are (anti-)periodic along $x,y$.  The background fields  $A^C$ and $A$ are smooth in  the northern  patch of $\spindle$ if $\alpha_2 = 1/\nN$, which is the value making the components $A^C_\varphi$ and $A_\varphi$ vanish at $\theta=0$. Analogously,  $A^C$ and $A$ exhibit  smoothness in  the southern  patch of $\spindle$ if $\alpha_2 = - 1/\nS$. Out of $\zeta, \widetilde \zeta$ and their Hermitean conjugate the bilinears
\begin{align}
      K^\mu = \zeta \sigma^\mu \widetilde \zeta    ~ , \qquad   Y^\mu = \frac{ \zeta \sigma^\mu \widetilde \zeta^\dagger }{2 \,  \widetilde \zeta^\dagger \widetilde \zeta }    ~ , \qquad   \widetilde Y^\mu = - \frac{ \zeta^\dagger \sigma^\mu \widetilde \zeta }{2 \,   \zeta^\dagger \zeta }    ~ , \qquad   \widetilde K^\mu = \frac{ \zeta^\dagger  \sigma^\mu \widetilde \zeta^\dagger }{4 \, (\zeta^\dagger \zeta) ( \widetilde \zeta^\dagger \widetilde \zeta) }    ~ ,
\end{align}
can be built, which allows to rewrite the supersymmetry transformation as a cohomological complex \cite{Closset:2013sxa,Longhi:2019hdh}. For example, the Killing vector $K^\mu = \zeta \sigma^\mu \widetilde \zeta$ reads
\begin{align}
    & K = K^\mu \partial_\mu   =  k_0 \p{ \amf\partial_\varphi - \tau \partial_x + \partial_y } ~ , & k_0 = \frac{\im \snorm^2}{L \beta \tau_2} ~ ,
\end{align}
while $Y = Y^\mu \partial_\mu$ is
\begin{align}
    Y = - \p{2 L}^{-1} e^{\im \p{\alpha_2 \varphi + \alpha_3 x  + \alpha_4 y  } }\comm{  f^{-1} \partial_\theta + \im \p{\csc \theta} \partial_\varphi  }   ~ .
\end{align}
The Lie derivatives of the Killing spinors along $K$ are
\begin{align}
    &  \mathcal L_K \zeta = \im \Phi_R \zeta  ~ , & \mathcal L_K \widetilde \zeta = - \im \Phi_R \widetilde \zeta  ~ ,
\end{align}
where
\begin{align}
    \Phi_R = \iota_K A = \frac{k_0}{2} \p{ \amf\alpha_2 - \tau \alpha_3 + \alpha_4 }   ~ .
\end{align}


\subsection{Supersymmetry on $\blens\times S^1$} 

  We now derive the $\mathcal N=1$  rigid supergravity background on   twisted $\blens\times S^1$. On $\blens\times S^1$ we employ the line element
\begin{align}\label{eq: leblenss1}
    \dd s^2 = L^2\comm{  \fs^2 \dd \vartheta^2 + b_1^2 \sin^2\vartheta^2\p{ \tS \dd\varphi + \frac{\nS \dd\psi }{n} }^2 + b_2^2 \cos^2\vartheta^2\p{ \tN \dd\varphi + \frac{\nN \dd\psi }{n} }^2 + \beta^2 \dd t^2 } ~ ,
\end{align}
 with  $\vartheta\in\comm{0, \pi/2}$ and $\varphi, \psi, t\in[0, 2\pi)$. The parameter $\beta$ is the radius of $S^1$;  the integers  $n,   \tN, \tS , \nN , \nS$  fulfil $n\geq1$, $\gcd\p{\nN, \nS}=1$  and $\p{\nN \tS - \nS \tN } = 1 $; while $b_1, b_2$ are squashing parameters that  induce a refinement of the partition function and the  squashing  function $\fs=\fs(\vartheta)$  satisfies $ \fs(0) = b_1 $ and  $ \fs(\pi/2) = b_2$. The line element (\ref{eq: leblenss1}) can be rewritten as  
 \begin{align}
    \dd s^2 & =  L^2\comm{  \fs^2 \dd \vartheta^2  + h_{11} \dd\varphi^2 + h_{22}\p{ A^{(1)} + \dd\psi }^2 + \beta^2 \dd t^2 }   ~ , \nn\\
    h_{11} & = \frac{b_1^2 b_2^2 \sin^2\p{2\vartheta}}{4\p{ \nS^2 b_1^2 \sin^2\vartheta  + \nN^2 b_2^2 \cos^2\vartheta }} ~ , \nn\\
    h_{22} & =  n^{-2}\p{ \nN^2 b_2^2 \cos^2\vartheta + \nS^2 b_1^2 \sin^2\vartheta} ~ , \nn\\
    A^{(1)} & =  \frac{n\p{ \nN \tN b_2^2 \cos^2\vartheta + \nS \tS b_1^2 \sin^2\vartheta }}{ \nN^2 b_2^2 \cos^2\vartheta + \nS^2 b_1^2 \sin^2\vartheta} \dd\varphi ~ ,  
\end{align}
with
\begin{align}
	& \dd s^2|_\spindle = L^2\p{  \fs^2 \dd \vartheta^2  + h_{11} \dd\varphi^2 } ~ , & \frac1{4\pi} \int_\spindle \dd\vartheta \dd\varphi \sqrt{g_\spindle}R_\spindle= \frac{1}{\nS} +  \frac{1}{\nN} ~ , 
\end{align}
 manifestly expressing  $\blens$ as a non-trivial circle fibration over a spindle $\spindle$, with flux
\begin{align}
	  \frac1{2\pi} \int_\spindle \dd A^{(1)} = \frac{n \, \tS}{\nS} - \frac{n \, \tN}{\nN} = \frac{n}{\nN \nS} ~ .
\end{align}
The branched lens space $\blens$ displays genuine orbifold singularities whenever $\gcd\p{n \, \tN, \nN}\neq1$ or $\gcd\p{n \, \tS, \nS}\neq1$ \cite{Inglese:2023tyc}~. We adopt the frame
\begin{align}
    \ee^1 & = L \, \fs ~ , \nn\\
    \ee^2 & = L \,  b_1 \sin\theta  \p{ \tS \dd\varphi + \frac{\nS \dd\psi}{n} } ~ , \nn\\
     \ee^3 & = L \,  b_2 \cos\theta  \p{ \tN \dd\varphi + \frac{\nN \dd\psi}{n} }  ~ , \nn\\ 
     \ee^4 & = L \, \beta \,   \dd t  ~ ,
\end{align}
where the spinors
\begin{align}
     \zeta_\alpha  & =  \snorm e^{\frac{\im}{2}\p{\alpha_2  \varphi  + \alpha_3 \psi  + \alpha_4  t }} \begin{pmatrix} -\im\cos\p{\vartheta/2} \\  - \sin\p{\vartheta/2} \end{pmatrix}_\alpha    ~ , \nn\\
      \widetilde \zeta^{\dot\alpha} & =  \snorm e^{-\frac{\im}{2}\p{ \alpha_2  \varphi  + \alpha_3 \psi  + \alpha_4  t  }} \begin{pmatrix} -\im\sin\p{\vartheta/2} \\   \cos\p{\vartheta/2} \end{pmatrix}^{\dot\alpha}    ~ ,
\end{align}
solve both the Killing and the Conformal Killing spinor equations  with background fields
\begin{align}
    A^C & = \p{ \alpha_2 +  \frac{ \tN b_2 - \tS b_1}{\fs}    }\frac{\dd\varphi}{2} + \p{ \alpha_3 +  \frac{ \nN b_2 - \nS b_1}{\fs}    }\frac{\dd\psi}{2}  + \p{ \alpha_4 -  \frac{ \im \beta }{\fs}    }\frac{\dd t}{2}  ~ , \nn\\
    V  & = \frac{b_1 \sin^2\vartheta\p{n \tS \dd\varphi + \nS \dd\psi } - b_2 \cos^2\vartheta\p{n \tN \dd\varphi + \nN \dd\psi }}{n \, \fs} + \kappa \, K^\flat   ~ , \nn\\
    A  & =  A^C + \frac{3}{2}V  ~ ,
\end{align}
where $K^\flat$ is the 1-form dual to the Killing vector $K^\mu\partial_\mu = \p{\zeta \sigma^\mu \widetilde \zeta   }\partial_\mu$, which  reads
\begin{align}
	    & K = K^\mu \partial_\mu   =  \knorm \comm{ \im \beta \p{\frac{\nN}{b_1} + \frac{\nS}{b_2} }\partial_\varphi - \im \, n \beta \p{ \frac{\tN}{b_1} + \frac{\tS}{b_2}  }\partial_\psi +\partial_t } ~ ,  & \knorm = \frac{\snorm^2}{L\beta}  ~ .
\end{align}
Furthermore, 
\begin{align}
	     Y & = Y^\mu \partial_\mu  = e^{\im \p{ \alpha_2  \varphi  + \alpha_3 \psi  + \alpha_4  t  }}\left\{ \frac{\im \partial_\vartheta }{2 L \fs}  \right. \nn\\
	     &  - \frac{\nN b_1^{-1} \cot\vartheta  - \nS b_1^{-1} \tan\vartheta }{2 L}\partial_\varphi + \left. \frac{n\p{\tN b_1^{-1} \cot\vartheta  - \tS b_1^{-1} \tan\vartheta }}{2 L} \partial_\psi  \right\}  ~ .
\end{align}
If $\alpha_4\in \mathbb Z$, then $\zeta$ and $\widetilde \zeta$ are (anti-)periodic along $t$.  The 1-forms  $A^C$ and $A$ are smooth on $\blens \times S^1$ if $\alpha_2 = \tS - \tN $ and $\alpha_3 = \p{\nS - \nN}/n $, giving 
\begin{align}
	\Phi_R = \iota_K A = \frac{\knorm}{2}\p{ \frac{\im \beta}{b_1} + \frac{\im \beta}{b_2} + \alpha_4} ~ .
\end{align}


\section{Localization}\label{sec: susylocalization}



In this section we perform   localization for vector and chiral multiplets on $\spindle \times T^2$ and $\blens \times S^1$. In general, $\mathcal N=1$  vector multiplets contain  a gauge field  $a_\mu $, chiral gauginos  $\lambda_\alpha , \widetilde \lambda^{\dot \alpha} $ and an auxiliary field $D$, all in the adjoint representation of a gauge group $G$. The R-charges of the tuple $\p{a_\mu , \lambda, \widetilde \lambda , D} \in {\rm adj}_G$ are  $\p{0, +1 , -1 , 0}$, respectively, with ${\rm adj}_G$ being the adjoint
representation of the gauge group $G$. On the other hand, matter fields are encoded by $\mathcal N=1$  chiral multiplets   , which consist  of a  complex scalar  $\phi  $, a   Weyl spinor $\psi_\alpha $ and  an auxiliary field  $F$ with R-charges $\p{r , r - 1  , r - 2 }$  and transforming in a representation $\mathfrak R_G$ of   $G$. We use the definitions and conventions  reported in \cite{Iannotti:2023jji}, where  supersymmetry variations, covariant derivatives, reality conditions, Lagrangians  and cohomological fields for   $\mathcal N=1$ vector and chiral multiplets on an arbitrary  supersymmetric background were reported.

We now compute the partition function of gauge theories coupled to matter via supersymmetric localization \cite{Pestun:2007rz}. We focus on Abelian gauge theories as the generalization to the non-Abelian case is straightforward. We start by deriving the supersymmetric locus solving the BPS equations; then, we will compute the 1-loop determinant of the fluctuations over the BPS locus.

 \subsection{Four-Dimensional Spindle Index}

\paragraph{BPS locus.} 

  On the BPS locus we have $\lambda|_{\rm BPS} = \widetilde\lambda|_{\rm BPS}=0$  and $\delta \lambda|_{\rm BPS} = \delta \widetilde \lambda|_{\rm BPS} = 0$. Let 
\begin{align}
    a = \p{ a_\varphi  + \frac{\beta_2}{2} }\dd \varphi + \p{ a_x   + \frac{\beta_3}{2} }\dd x  + \p{ a_y   + \frac{\beta_4}{2} }\dd y  ~ ,
\end{align}
be the ansatz for the BPS gauge field $ a_\mu$ of the  vector multiplet, with $a_{\phi, x , y}$ being functions of $\theta$ only and $\beta_{2,3,4}$ being constant flat connections. We impose
\begin{align}
    & a_\varphi\p{0} = \frac{\mN}{\nN}   ~ , & a_\varphi\p{\pi } = \frac{\mS}{\nS}   ~ ,
\end{align}
so that the gauge flux   through $\spindle$ is
\begin{align}
    & \mathfrak f_G = \frac{1}{2 \pi }\int_\spindle \dd  a = \frac{\mS}{\nS} - \frac{\mN}{\nN} = \frac{\mathfrak m}{\nN \nS}  ~ , &  \mathfrak m = \nN \mS - \nS \mN  ~ ,
\end{align}
Thus,  $a|_\spindle $ represents   a $\mathcal O(-\mathfrak m) = \mathcal O(\nS \mN  - \nN \mS )$ orbibundle. The BPS equations for $a$ read
\begin{align}
     \iota_K \p{\dd a} = \mathcal L_K a - \dd \p{ \iota_K a}    = 0   ~ ,
\end{align}
where $\mathcal L_K$ is the Lie derivative along the  vector $K$. The equation above immediately implies
\begin{align}
      a_y = a_0 - \amf a_\varphi + \tau  a_x   ~ ,
\end{align}
where $a_0$ is a constant. Then, the BPS value of $ \iota_K a$ is
\begin{align}
      \Phi_G = \iota_K a|_{\rm BPS} = \frac{k_0}{2} \p{ \amf\beta_2 - \tau \beta_3 + \beta_4 + 2 a_0 }    ~ .
\end{align}
The other BPS equations determine the form of the auxiliary field $D$:
\begin{align}
       D|_{\rm BPS} = 2 \im Y^\mu \widetilde Y^\nu f_{\mu \nu} |_{\rm BPS} = \frac{ \csc\theta  }{L^2 f}  a'_\varphi(\theta)    ~ .
\end{align}
As for the chiral-multiplet   BPS locus, the vanishing of the supersymmetry variations $\psi|_{\rm BPS} = \widetilde\psi|_{\rm BPS}= \delta \psi|_{\rm BPS} = \delta \widetilde \psi|_{\rm BPS} = 0$ implies    $\phi|_{\rm BPS} = \widetilde \phi|_{\rm BPS} = 0 $  for arbitrary values of  $\Phi_{R,G}$.


\paragraph{One-loop determinant.}

 We mostly deal with 1-loop determinants for chiral multiplets as those for the vector multiplet can be obtained from that of an adjoint chiral with R-charge $r=2$. The 1-loop determinant is given by
\begin{align}\label{eq: 1Lunpairedeigenvaluesmaster}
        Z_{\text{1-L}}  = \frac{\det_{\ker L_Y} \delta^2 }{\det_{\ker L_{\widetilde Y} }  \delta^2 }   ~ ,
\end{align}
where the differential operators  $L_Y = Y^\mu D_\mu$, $L_{\widetilde Y} = \widetilde Y^\mu D_\mu$ are the contractions of the spinor bilinears $Y$ and $\widetilde Y$ with the covariant derivative $D_\mu$. Furthermore, 
\begin{align}
         \delta^2 = 2 \im \p{ \mathcal L_K - \im \widehat q_R \Phi_R - \im \widehat q_G \Phi_G }  ~ ,
\end{align}
with $\widehat q_{R,G}$ representing the charge  operators for R-symmetry and gauge/flavor-symmetry, respectively, acting on a field $X$ as $\widehat q_{R,G} X = q^X_{R,G} X$.  Eigenfunctions of $\delta^2$ in the kernel of $L_{\widetilde Y}$ have the form
\begin{align}
        \Phi  = e^{\im \p{ n_\varphi  \varphi  + n_x x + n_y y   }  } \Phi_0(\theta) ~ ,
\end{align}
with eigenvalue
\begin{align}
          \lambda_\Phi  = 2  \comm{  k_0 \p{ - \amf n_\varphi  + \tau n_x - n_y }+   q_R^\Phi  \Phi_R +  q_G \Phi_G }  ~ ,
\end{align}
and $\Phi_0(\theta)$ fixed by the differential equation
\begin{align}
        L_{\widetilde Y }\Phi  = 0 ~ .
\end{align}
The function $\Phi_0(\theta)$ exhibits regularity at the north pole of $\spindle$ if it remains non-divergent at $\theta=0$, a condition satisfied if
\begin{align}
        n_\varphi \leq  q_R^\Phi  \p{ \frac{\alpha_2}{2} - \frac{1}{2 \nN}  } + q_G \p{ \frac{\beta_2}{2} + \frac{\mN}{\nN}  }    ~ ,
\end{align}
while non-singularity of $\Phi_0(\theta)$ at the south  pole of $\spindle$ at $\theta = \pi $ requires  
\begin{align}
        n_\varphi \geq  q_R^\Phi  \p{ \frac{\alpha_2}{2} + \frac{1}{2 \nS}  } + q_G \p{ \frac{\beta_2}{2} + \frac{\mS}{\nS}  }    ~ .
\end{align}
The field $\Phi$ has a twisted periodicity along $\phi$ consistent with that of the Killing spinors if 
\begin{align}
        & n_\varphi  = n'_\varphi + q_R^\Phi  \frac{\alpha_2}{2}  ~ , & n'_\varphi  \in \mathbb Z  ~ .
\end{align}
Generalizing the latter to the gauge/flavour bundle yields 
\begin{align}
        & n_\varphi  = i_\varphi +  q_R^\Phi  \frac{\alpha_2}{2} +  q_G \frac{\beta_2}{2}  ~ , & i_\varphi  \in \mathbb Z  ~ .
\end{align}
and   
\begin{align}
          \cf{ \frac{\p{q_R^\Phi/2} + q_G \mS}{\nS} } \leq i_\varphi \leq \ff{ \frac{-\p{q_R^\Phi/2} + q_G \mN}{\nN} }       ~ ,
\end{align}
where $\ff \bullet$ and $\cf \bullet$ are the floor and the ceiling of $\bullet$, respectively. Regularity imposes no  constraints on the Fourier modes on $T^2$, $n_x , n_y$, which we take as integers: $n_x , n_y \in \mathbb Z$. Given $q_G\in \mathbb Z$ and $q_R^\Phi = r$ for a chiral-multiplet scalar, it is convenient to define the collective  integers   
 \begin{align}
          & \pN = q_G \mN - \st \p{r/2}      ~ , & \pS = q_G \mS +  \p{r/2}      ~ ,
\end{align}
giving 
\begin{align}
          - \ff{ - \pS/\nS } \leq i_\varphi \leq \ff{ \pN/\nN }       ~ .
\end{align}
If
\begin{align}
           \ff{ \pN/\nN } + \ff{ - \pS/\nS }  \geq 0      ~ ,
\end{align}
then the modes $\Phi$ of R-charge $r$ contribute to $Z_{\text{1-L}}$ by
          %
\begin{align}
          Z_\Phi  & = \prod_{i_\varphi = - \ff{- \pS/\nS}}^{\ff{\pN/\nN}}  \prod_{n_x , n_y \in \mathbb Z} \frac1{    - \amf i_\varphi  + \tau n_x - n_y +   \frac{r}{2}  \p{ - \tau \alpha_3 + \alpha_4}  +  q_G \p{ a_0 - \tau \frac{\beta_3}{2} + \frac{\beta_4}{2}} }  ~ .
\end{align}
By employing the linear change of variables
\begin{align}
     & i_\varphi = - h_\varphi + \ff{\pN/\nN} ~ , & 0 \leq h_\varphi \leq \mathfrak b - 1  ~ ,
\end{align}
where
\begin{align}\label{eq: frakb}
   & \mathfrak b = 1 + \st \ff{ \st \frac{ \mathfrak m \, \tN - \p{r/2} \tN }{\nN}} + \ff{ - \frac{ \mathfrak m \, \tS + \p{r/2} \tS }{\nS}}  ~ , &  \nN \tS - \nS \tN = 1 ~ , \qquad t_\pm \in \mathbb Z ~ ,
\end{align}
was introduced in \cite{Inglese:2023wky,Inglese:2023tyc}. We find
\begin{align}
          Z_\Phi  & = \prod_{h_\varphi =0 }^{ \mathfrak b - 1 }  \prod_{n_x , n_y \in \mathbb Z} \comm{  \amf \p{ h_\varphi -  \ff{\frac{\pN}{\nN}} } + \tau n_x - n_y +   \frac{r}{2}  \p{ - \tau \alpha_3 + \alpha_4}  +  q_G \p{ a_0 - \tau \frac{\beta_3}{2} + \frac{\beta_4}{2}} }^{-1}  ~ , \nn\\
          %
          %
          %
          & = \prod_{h_\varphi =0 }^{ \mathfrak b - 1 }  \prod_{n_x , n_y \in \mathbb Z} \comm{  \amf\p{ h_\varphi + q_i^\Phi  \frac{\mathfrak f_i}{2 }   + \frac{\rem{\pN}{\nN}}{\nN} } + \tau n_x - n_y - q_i^\Phi  \gamma_i }^{-1}  ~ ,
\end{align}
with $\rem{\bullet}{\diamond}$ being the reminder of the integer division of  $\bullet$ by $\diamond$ and  
\begin{align}\label{eq: chemicalpotentialsspindlet2tt}
    q_i^\Phi \gamma_i & =  q_R^\Phi \gamma_R + q_G \gamma_G =  r  \gamma_R + q_G \gamma_G ~ , \nn\\
    \gamma_R & =    \amf   \frac{\chi_-}{4} + \tau \frac{ \alpha_3}{2} - \frac{\alpha_4}{2} ~ , \nn\\
    \gamma_G & =      \frac{\amf }{2} \p{ \frac{\mS}{ \nS} +   \frac{\mN}{ \nN} } + \tau \frac{\beta_3}{2} - \frac{\beta_4}{2} - a_0  ~ .
\end{align}
As
\begin{align}
    & q_i^\Phi  \frac{\mathfrak f_i}{2 }   + \frac{\rem{\pN}{\nN}}{\nN}  =  \frac{1}{2}\p{1 - \mathfrak b - \mathfrak c }   ~ , & \mathfrak c  =   \frac{ \rem{ - \pS}{ \nS} }{\nS} -  \st \frac{ \rem{ \st \, \pN}{ \nN} }{\nN}  ~ , 
\end{align}
we can write 
\begin{align}
          Z_\Phi  &  = \prod_{h_\varphi =0 }^{ \mathfrak b - 1 }  \prod_{n_x , n_y \in \mathbb Z} \acomm{  \amf\comm{ h_\varphi +  \frac{1}{2}\p{1 - \mathfrak b - \mathfrak c } } + \tau n_x - n_y - q_i^\Phi  \gamma_i }^{-1}  ~ .
\end{align}
Regularizing the latter\footnote{See for instance Appendix A of \cite{Iannotti:2023jji}~.} yields 
\begin{align}\label{eq: ZCMbgeq1}
    Z|_{\mathfrak b \geq 1} & = e^{2 \pi \im \Psi  } \prod_{ j =0 }^{ \mathfrak b - 1 } \prod_{n\in \mathbb N}\frac{ 1 }{\p{ 1 - z^{-1}  q^{ j +   \frac{1}{2}\p{1 - \mathfrak b  }  }     p^n }  \p{ 1 -  z \,  q^{ - j -   \frac{1}{2}\p{1 - \mathfrak b  }  }    p^{n+1} } }   ~ , \nn\\
        \Psi & = - \frac{\mathfrak b}{24 \, \tau }\comm{2 + 12 \,  q_i^\Phi \gamma_i\p{ q_\ell^\Phi \gamma_\ell  + 1 + \tau + \amf \mathfrak c } + 2 \tau\p{3 + \tau} + 6 \amf \mathfrak c \p{1 + \tau } + \amf^2\p{ \mathfrak b^2 - 1 + 3 \mathfrak c^2}}  ~ , \nn\\
          p & = e^{2 \pi \im \tau }   ~ , \qquad q = e^{2 \pi \im \amf  }   ~ , \qquad z = q^{  \mathfrak c/2} e^{  2 \pi \im q_i^\Phi \gamma_i  }   ~ .
\end{align}
Analogously, eigenfunctions of $\delta^2$ in the kernel of $L_{ Y}$ have the form
\begin{align}
        B  = e^{\im \p{ \ell_\varphi  \varphi  + \ell_x x + \ell_y y   }  } B_0(\theta) ~ ,
\end{align}
with eigenvalue
\begin{align}
          \lambda_B  = 2  \comm{  k_0 \p{ - \amf\ell_\varphi  + \tau \ell_x - \ell_y }+   q_R^B  \Phi_R +  q_G \Phi_G }  ~ ,
\end{align}
and $B_0(\theta)$ fixed by the differential equation
\begin{align}
        L_{ Y } B  = 0 ~ ,
\end{align}
Non-singularity of $B_0(\theta)$ at the north pole of $\spindle$ at $\theta = 0$ requires  
\begin{align}
        \ell_\varphi \geq  q_R^B  \p{ \frac{\alpha_2}{2} - \frac{1}{2 \nN}  } + q_G \p{ \frac{\beta_2}{2} + \frac{\mN}{\nN}  }    ~ ,
\end{align}
while non-singularity of $B_0(\theta)$ at the south  pole of $\spindle$ at $\theta = \pi $ requires  
\begin{align}
        \ell_\varphi \leq  q_R^B  \p{ \frac{\alpha_2}{2} + \frac{1}{2 \nS}  } + q_G \p{ \frac{\beta_2}{2} + \frac{\mS}{\nS}  }    ~ .
\end{align}
The field $B$ has a twisted periodicity along $\phi$ consistent with those of the Killing spinors and of $\Phi$ if 
\begin{align}
        & \ell_\varphi  = j_\varphi + q_R^B \frac{\alpha_2}{2} + q_G \frac{\beta_2}{2}  ~ , & j_\varphi  \in \mathbb Z  ~ .
\end{align}
giving    
\begin{align}
           \cf{ \frac{-\p{q_R^B/2} + q_G \mN}{\nN} } \leq  j_\varphi \leq  \ff{ \frac{\p{q_R^B/2} + q_G \mS}{\nS} }    ~ ,
\end{align}
namely, for a Grassmann-odd scalar $B$ of R-charge $q_R^B = \p{r-2}$,
\begin{align}
           1 + \ff{ \frac{  \pN}{\nN} } = \cf{ \frac{ 1 + \pN}{\nN} } \leq  j_\varphi \leq  \ff{ \frac{ \pS - 1}{\nS} } = - 1 - \ff{ - \frac{ \pS }{\nS} }   ~ .
\end{align}
Regularity imposes no  constraints on the Fourier modes on $T^2$,  $\ell_x , \ell_y$, which we take as integers: $\ell_x , \ell_y \in \mathbb Z$. If
\begin{align}
          & \ff{ \frac{  \pN}{\nN} } +  \ff{ - \frac{ \pS }{\nS} }    \leq - 2    ~ , & \mathfrak b   \leq - 1    ~ ,
\end{align}
then the modes $B$ contribute to $Z_{\text{1-L}}$ by
\begin{align}
          Z_B  & = \prod_{j_\varphi = \cf{\frac{1 + \pN}{\nN}}}^{\ff{\frac{\pS-1}{\nS}}}  \prod_{\ell_x , \ell_y \in \mathbb Z} \comm{    - \amf j_\varphi  + \tau n_x - n_y +   \frac{r-2}{2}  \p{ - \tau \alpha_3 + \alpha_4}  +  q_G \p{ a_0 - \tau \frac{\beta_3}{2} + \frac{\beta_4}{2}} }  ~ .
\end{align}
After the linear change of variables
\begin{align}
     & j_\varphi = -  k_\varphi + \ff{\p{\pS-1}/\nS} ~ , & 0 \leq k_\varphi \leq - \mathfrak b - 1   ~ ,
\end{align}
the 1-loop determinant reads  
\begin{align}
          Z_B 
          %
          %
          & = \prod_{k_\varphi = 0 }^{ - \mathfrak b - 1}  \prod_{\ell_x , \ell_y \in \mathbb Z} \comm{     \amf  \p{ k_\varphi - q_i^B \frac{\mathfrak f_i}{2 } + \frac{ \rem{\pS-1}{ \nS} }{\nS}   }  + \tau \ell_x - \ell_y - q_i^B \gamma_i  }  ~ .
\end{align}
Since 
\begin{align}
           - q_i^B \frac{\mathfrak f_i}{2 } + \frac{ \rem{\pS-1}{ \nS} }{\nS}   
          %
          %
          %
          %
          & =  \frac{1}{2} \p{ 1 + \mathfrak b - \mathfrak c - \chi_-  } ~ , 
\end{align}
we can write
\begin{align}
          Z_B & = \prod_{k_\varphi = 0 }^{ - \mathfrak b - 1}  \prod_{\ell_x' , \ell_y' \in \mathbb Z} \acomm{     \amf  \comm{ k_\varphi  + \frac{1}{2} \p{ 1 + \mathfrak b - \mathfrak c   }   }  + \tau \ell_x' - \ell_y' - q_i^\Phi \gamma_i  }  ~ .
\end{align}
After regularization  we obtain 
\begin{align}\label{eq: ZCMbleqm1}
    Z|_{\mathfrak b\leq -1} & = e^{2 \pi \im \Psi }  \prod_{j = 0 }^{ - \mathfrak b - 1}   \prod_{n\in\mathbb N}\p{  1 - z^{-1} q^{ j  + \frac{1}{2} \p{ 1 + \mathfrak b     }    }  p^{n} }  \p{ 1 -  z \, q^{ - j  - \frac{1}{2} \p{ 1 + \mathfrak b     }    }   p^{n+1} }  ~ ,   
\end{align}
where $\Psi$ is reported in (\ref{eq: ZCMbgeq1}). Finally, neither $\Phi$ nor $B$ contributes to $Z_{\text{1-L}}$ if $\mathfrak b=0$, hence  $Z|_{\mathfrak b = 0} = 1$.


\paragraph{Index  theorem on orbifolds.}

The 1-loop determinant on $\spindle \times T^2$ can also be obtained from the equivariant index of the operator pairing the  fundamental fields  in the cohomological complex \cite{Pestun:2007rz}. In the conventions of \cite{Closset:2013sxa} such  an operator is $L_{\widetilde Y}$ and the index is calculated  with respect to the equivariant action
 \begin{align}
         & g = \exp\p{ - \im \epsilon \delta^2 } = g_{\spindle} \,  g_{T^2} ~ , & g_\spindle  = \exp\p{ - \im \epsilon \delta^2|_\spindle  }  ~ , \qquad g_{T^2}  = \exp\p{ - \im \epsilon \delta^2|_{T^2}  }  ~ ,
\end{align}
with 
\begin{align}
    \delta^2|_\spindle  &  =  2 \im k_0 \, \amf  \p{  \partial_\varphi - \im q_R^\Phi \frac{\alpha_2}{2} - \im q_G  \frac{\beta_2}{2}  } ~ , \nn\\
    \delta^2|_{T^2} & =  2 \im k_0 \comm{ - \tau \partial_x + \partial_y - \im q_R^\Phi \p{ - \tau \frac{\alpha_3}{2} + \frac{\alpha_4}{2} } - \im q_G \p{ a_0 - \tau \frac{\beta_3}{2} + \frac{\beta_4}{2} }  } ~ ,
\end{align}
where the flat connections on $\spindle$ are those making the R-symmetry connection $A$ and the gauge field  $a$ smooth in the northern and southern  patches $\mathcal U_\pm$:
\begin{align}
    & \alpha_2|_{\mathcal U_\pm}    = \pm 1/n_\pm  ~ , & \beta_2|_{\mathcal U_\pm}   = - 2 \, \mathfrak m_\pm/n_\pm ~ ,
\end{align}
giving, on a field $\Phi^{(r)}$ of R-charge $r$,
\begin{align}
    & g|_{\mathcal U_+}   =  e^{ 2  \epsilon  k_0 \, \amf  \p{  \partial_\varphi + \im   \frac{\pN}{\nN}  } } ~ , & g|_{\mathcal U_-}   = e^{   2 \epsilon  k_0 \, \amf  \p{  \partial_\varphi +   \im   \frac{\pS}{\nS}  } } ~ .
\end{align}
The operator $g_{T^2}$ acts freely, whereas $g_\spindle$ has fixed points at the north and the south pole of the spindle. The tangent space $\mathbb C/\mathbb Z_{\nN}$ in  a neighbourhood of $\theta=0$ is parametrized by a complex coordinate $\zN$ satisfying 
\begin{align}
     & \zN =   \theta \exp\p{\im \varphi / \nN} ~ , & \zN \sim \wN \zN ~ ,  
\end{align}
while the coordinate on  the tangent space $\mathbb C/\mathbb Z_{\nS}$ in  a neighbourhood of $\theta=\pi$ fulfils    
\begin{align}
    & \zS =   \p{\pi - \theta} \exp\p{ - \im \varphi / \nS }  ~ , & \zS \sim \wS^{-1} \zS ~ .
\end{align}
In such coordinates, the operator $L_{\widetilde Y}$ in the northern and southern neighbourhoods  $\mathcal U_\pm$ of $\spindle$ reads
\begin{align}
    & L_{\widetilde Y}|_{\mathcal U_+} = - \p{\nN \, L}^{-1} \partial_+    ~ , & L_{\widetilde Y}|_{\mathcal U_-} =  \p{\nS \, L}^{-1} \partial_-     ~ ,
\end{align}
where $\partial_\pm = \partial_{z_\pm}$. The equivariant action on coordinates in the northern patch of the spindle is 
\begin{align}
    & g|_{\mathcal U_+} \circ  \zN =   \qN  \zN ~ , & \qN =  e^{ -  2 \im  \epsilon  k_0 \, \amf / \nN  }  ~ , \qquad \mathfrak q = \qN^{\nN} =  e^{ -  2 \im  \epsilon  k_0 \, \amf  }  ~ , 
\end{align}
while the equivariant action   on fields reads 
\begin{align}
     g|_{\mathcal U_+} \circ  \Phi^{(r)} =  \qN^{ -   \pN  }  \Phi^{(r)}    ~ ,
\end{align}
and on 1-forms as
\begin{align}
    g|_{\mathcal U_+} \circ  \partial_+ \Phi^{(r)} =  \qN^{ - 1 -   \pN  }  \partial_+ \Phi^{(r)}    ~ . 
\end{align}
The north-pole contribution to the $g$-equivariant index of $L_{\widetilde Y}$ is then obtained by   applying the  orbifold projection  used in  \cite{MEINRENKEN1998240,Inglese:2023wky,Inglese:2023tyc} to
\begin{align}
    I_{+} =  \frac{ \qN^{  -   \pN  }  - \qN^{ - 1 -   \pN  }  }{\p{1 - \qN}\p{1 - \qN^{-1}} } =  \frac{ \qN^{  -   \pN  }   }{1 - \qN } ~ .
\end{align}
This projection operates through the substitution $q_\pm\to q_\pm w_\pm^j$ followed by averaging over $j=0, \dots, \p{n_\pm-1}$, which results  in
\begin{align}
    I_{ \mathcal U_+}  = \frac{1}{\nN} \sum_{j=0}^{\nN - 1}  \frac{ \wN^{ - j \,  \pN} \qN^{  -   \pN  }   }{1 - \wN^j \qN } =    \frac{   \mathfrak q^{ \p{ \rem{\pN}{ \nN} - \pN  } /\nN}   }{1 -  \mathfrak q } =    \frac{   \mathfrak q^{ - \ff{\pN/\nN}}   }{1 -  \mathfrak q } ~ . 
\end{align}
Including the free action $g_{T^2}$ provides
\begin{align}
    I_{ \mathcal U_+ \times T^2} & = \sum_{n_x , n_y \in \mathbb Z} e^{ - 2 \im  \epsilon  k_0 \comm{ \tau n_x - n_y + r \p{ - \tau \frac{\alpha_3}{2} + \frac{\alpha_4}{2} }  + q_G \p{ a_0 - \tau \frac{\beta_3}{2} + \frac{\beta_4}{2} }  } }    \frac{   \mathfrak q^{ - \ff{\pN/\nN}}   }{1 -  \mathfrak q } ~ , \nn\\ 
     & = \sum_{n_x , n_y \in \mathbb Z} \sum_{\ell \in \mathbb N} e^{ - 2 \im  \epsilon  k_0 \comm{ \omega \p{  \ell  - \ff{\pN/\nN} } + \tau n_x - n_y + r \p{ - \tau \frac{\alpha_3}{2} + \frac{\alpha_4}{2} }  + q_G \p{ a_0 - \tau \frac{\beta_3}{2} + \frac{\beta_4}{2} }  } }    ~ . 
\end{align}
As in e.g. (5.18) of \cite{Assel:2016pgi}, $I_{ \mathcal U_+ \times T^2}$ translates   into an infinite product as follows:
\begin{align}
    Z_+ & = \prod_{n_x , n_y \in \mathbb Z} \prod_{\ell \in \mathbb N} \comm{ \amf  \p{  \ell  - \ff{\frac{\pN}{\nN} } } + \tau n_x - n_y + \frac{r}{2} \p{ - \tau  \alpha_3  +  \alpha_4 }  + q_G \p{ a_0 - \tau \frac{\beta_3}{2} + \frac{\beta_4}{2} }  }^{-1}    ~ , \nn\\
    & = \prod_{n_x , n_y \in \mathbb Z} \prod_{\ell \in \mathbb N} \acomm{ \amf \comm{  \ell  + \frac{1}{2}\p{1 - \mathfrak b - \mathfrak c} } + \tau n_x - n_y - q_i^\Phi \gamma_i  }^{-1}    ~ ,
\end{align}
which after regularization  becomes
\begin{align}\label{eq: ZblockN}
           Z_+ & =  \frac{e^{ 2 \pi \im \Psi_+ } }{ \p{ z^{-1}  q^{ \frac{1}{2}\p{1 - \mathfrak b} } ; q , p }_\infty  \p{ z \, p \, q^{ - \frac{1}{2}\p{1 - \mathfrak b} }  ; q^{-1} , p }_\infty} =   e^{ 2 \pi \im \Psi_+}   \Gamma_e\p{ z^{-1}  q^{ \frac{1}{2}\p{1 - \mathfrak b} } ; q , p }  ~ , \nn\\ 
          \Psi_+ & =  - \frac{1}{48 \tau \amf  }\comm{ 1 + 2 q_i^\Phi \gamma_i + \tau + \amf \p{\mathfrak c + \mathfrak b} } \nn\\
          & \times \acomm{ 2 \tau + 4 q_i^\Phi \gamma_i\p{ 1 + q_i^\Phi \gamma_i + \tau  } + 2 \amf \p{\mathfrak c + \mathfrak b}\p{1 + 2 q_i^\Phi \gamma_i + \tau} + \amf^2\comm{ \p{\mathfrak c + \mathfrak b}^2 - 1 } } ~ . 
\end{align}
On the other hand, the equivariant action acts on coordinates in the southern patch of $\spindle$ as
\begin{align}
    & g|_{\mathcal U_-} \circ  \zS =   \qS^{-1} \zS ~ , & \qS =  e^{  - 2 \im  \epsilon  k_0 \, \amf / \nS  }  ~ , \qquad  \qS^{\nS} =  \mathfrak q  ~ , 
\end{align}
while the equivariant action on fields reads
\begin{align}
     g|_{\mathcal U_-} \circ  \Phi^{(r)} =  \qS^{ -   \pS  }  \Phi^{(r)}    ~ , 
\end{align}
and the equivariant action on 1-forms is
\begin{align}
    g|_{\mathcal U_-} \circ  \partial_- \Phi^{(r)} =  \qS^{  1 -   \pS  }  \partial_- \Phi^{(r)}    ~ .
\end{align}
Thus, given the auxiliary quantity
\begin{align}
    I_{-} =  \frac{ \qS^{  -   \pS  }  - \qS^{ 1 -   \pS  }  }{\p{1 - \qS}\p{1 - \qS^{-1}} } =  \frac{ \qS^{  -   \pS  }   }{1 - \qS^{-1} } ~ , 
\end{align}
the south-pole contribution to the index turns out to be
\begin{align}\label{eq: indexLYtildesouthpole}
    I_{ \mathcal U_-}  = \frac{1}{\nS} \sum_{j=0}^{\nS - 1}  \frac{ \wS^{ - j \,  \pS} \qS^{  -   \pS  }   }{1 - \wS^{-j} \qS^{-1} } =    \frac{   \mathfrak q^{  \p{ - \rem{ - \pS}{\nN} - \pS  } /\nS}   }{1 -  \mathfrak q^{-1} } =    \frac{   \mathfrak q^{  \ff{ - \pS/\nS}}   }{1 -  \mathfrak q^{-1} } = -    \frac{   \mathfrak q^{ 1 +  \ff{ - \pS/\nS}}   }{1 -  \mathfrak q } ~ , 
\end{align}
whose uplift to $\spindle \times T^2$ reads
\begin{align}
    I_{ \mathcal U_- \times T^2} & = - \sum_{\ell_x , \ell_y \in \mathbb Z} e^{ - 2 \im  \epsilon  k_0 \comm{ \tau \ell_x - \ell_y + r \p{ - \tau \frac{\alpha_3}{2} + \frac{\alpha_4}{2} }  + q_G \p{ a_0 - \tau \frac{\beta_3}{2} + \frac{\beta_4}{2} }  } }  \frac{   \mathfrak q^{ 1 +  \ff{ - \pS/\nS}}   }{1 -  \mathfrak q  } ~ , \nn\\ 
     & = - \sum_{\ell_x , \ell_y \in \mathbb Z} \sum_{ j \in \mathbb N} e^{ - 2 \im  \epsilon  k_0 \comm{ \omega \p{   j + 1 +  \ff{ - \frac{\pS}{\nS}} } + \tau \ell_x - \ell_y + r \p{ - \tau \frac{\alpha_3}{2} + \frac{\alpha_4}{2} }  + q_G \p{ a_0 - \tau \frac{\beta_3}{2} + \frac{\beta_4}{2} }  } }    ~ .
\end{align}
The infinite-product counterpart of $I_{ \mathcal U_- \times T^2}$ is
\begin{align}
    Z_- & = \prod_{\ell_x , \ell_y \in \mathbb Z} \prod_{j \in \mathbb N} \comm{ \omega \p{   j + 1 +  \ff{ - \frac{\pS}{\nS}} } + \tau \ell_x - \ell_y + \frac{r}{2} \p{ - \tau \alpha_3 + \alpha_4 }  + q_G \p{ a_0 - \tau \frac{\beta_3}{2} + \frac{\beta_4}{2} }  }    ~ , \nn\\
    %
    %
    & = \prod_{\ell'_x , \ell'_y \in \mathbb Z} \prod_{j \in \mathbb N} \acomm{ \omega \comm{   j +\frac{1}{2}\p{1 + \mathfrak b - \mathfrak c} } + \tau \ell'_x - \ell'_y - q_i^\Phi \gamma_i   }    ~ .
\end{align}
Regularizing the latter gives
\begin{align}\label{eq: ZblockS}
           Z_- & =  {e^{ 2  \pi \im  \Psi_-  } }  \p{ z^{-1} q^{\frac{1}{2}\p{ 1 + \mathfrak b  }} ; q , p   }_\infty  \p{ z \, p \, q^{-\frac{1}{2}\p{ 1 + \mathfrak b  }} ; q^{-1} , p }_\infty    = \frac{  {e^{ 2  \pi \im  \Psi_-  } }  } { \Gamma_e\p{ z^{-1} q^{\frac{1}{2}\p{ 1 + \mathfrak b  }} ; q , p   }_\infty  }  ~ , \nn\\
          \Psi_- & =  \frac{1}{48 \tau \amf  }\comm{1 + \tau + 2 q_i^\Phi \gamma_i + \amf\p{ \mathfrak c - \mathfrak b }   } \nn\\
          & \times \acomm{ 2 \tau + 4 q_i^\Phi \gamma_i\p{1 + q_i^\Phi \gamma_i +  \tau} + 2 \amf  \p{\mathfrak c - \mathfrak b}\p{1 + 2 q_i^\Phi \gamma_i + \tau } + \amf^2\comm{ \p{\mathfrak c - \mathfrak b}^2 - 1} } ~ .
\end{align}
The product of the blocks (\ref{eq: ZblockN})  and (\ref{eq: ZblockS}) provides
\begin{align}\label{eq: ZCMit}
    Z_{\text{1-L}} & = Z_+ Z_- = e^{2 \pi \im \Psi }  \frac{ \Gamma_e\p{ z^{-1}  q^{ \frac{1}{2}\p{1 - \mathfrak b} } ; q , p }  } { \Gamma_e\p{ z^{-1} q^{\frac{1}{2}\p{ 1 + \mathfrak b  }} ; q , p   }  } ~ , \nn\\
    \Psi & = \Psi_+ + \Psi_- ~ ,
\end{align}
where $\Psi$ is written in (\ref{eq: ZCMbgeq1}). The function $Z_{\text{1-L}}$ in (\ref{eq: ZCMit}), valid for any $\mathfrak b \in \mathbb Z$, satisfies $Z_{\text{1-L}}=1$ for $\mathfrak b=0$ and correctly matches $Z_\Phi$ and $Z_B$ for $\mathfrak b\geq 1$ and  $\mathfrak b\leq 1$, respectively. The 1-loop determinant above can be rewritten in terms of elliptic gamma functions, providing
\begin{align}\label{eq: ZCMitRG}
    Z_{\text{1-L}}^{\rm CM}  &   = \prod_{\rho \in \mathfrak R_G} e^{2 \pi \im \Psi\p{\rho\p{\mathfrak m} , \rho\p{\gamma_G} } } \frac{  \Gamma_e\p{ z^{-\rho} q^{\frac{1}{2}\comm{1 - \mathfrak b\p{\rho\p{\mathfrak m} } }} ; q , p   } }{ \Gamma_e\p{ z^{-\rho}  q^{ \frac{1}{2}\comm{1 + \mathfrak b\p{\rho\p{\mathfrak m} } } } ; q , p }}    ~ ,
\end{align}
for a chiral multiplet in the representation $\mathfrak R_G$. The one-loop determinant for a vector multiplet in the adjoint of the gauge group $G$ can be obtained from (\ref{eq: ZCMitRG}) by setting $r=2$ and  $ \mathfrak R_G = {\rm adj}_G$:
\begin{align}\label{eq: ZVMitG}
    Z_{\text{1-L}}^{\rm VM}  &   = \prod_{\alpha \in {\rm adj}_G} e^{2 \pi \im \Psi\p{\alpha\p{\mathfrak m} , \rho\p{\gamma_G} } } \left. \frac{  \Gamma_e\p{ z^{-\alpha} q^{\frac{1}{2}\comm{1 - \mathfrak b\p{\alpha\p{\mathfrak m} } }} ; q , p   } }{ \Gamma_e\p{ z^{-\alpha}  q^{ \frac{1}{2}\comm{1 + \mathfrak b\p{\alpha\p{\mathfrak m} } } } ; q , p }}\right|_{r=2}    ~ .
\end{align}



The orbifold index theorem  also allows us to formulate an educated guess for partition functions on $\mathcal N=1$ supersymmetric   $\spindle\times T^2$  with R-symmetry anti-twist on $\spindle$. Indeed, the equiviariant  index of the operator $L_{\widetilde Y}$ on anti-twisted  $\spindle$ is the sum of two contributions: one is the very $I_{ \mathcal U_-} $ reported in (\ref{eq: indexLYtildesouthpole}), while the other is \cite{Inglese:2023wky,Inglese:2023tyc}
\begin{align}
    \widetilde I_{ \mathcal U_+} =    \frac{   \mathfrak q^{ \ff{-\pN/\nN}}   }{1 -  \mathfrak q^{-1} } =  -   \frac{   \mathfrak q^{ 1 + \ff{-\pN/\nN}}   }{1 -  \mathfrak q }  ~ . 
\end{align}
We can uplift $\widetilde I_{ \mathcal U_+}$ to $\widetilde I_{ \mathcal U_+ \times T^2}$ by including  the free action on the torus given by  $g_{T^2}$, which is supposed to be independent of the R-symmetry twist present on $\spindle$. The result is
\begin{align}
    \widetilde I_{ \mathcal U_+ \times T^2} & = \sum_{n_x , n_y \in \mathbb Z} e^{ - 2 \im  \epsilon  k_0 \comm{ \tau n_x - n_y + r \p{ - \tau \frac{\alpha_3}{2} + \frac{\alpha_4}{2} }  + q_G \p{ a_0 - \tau \frac{\beta_3}{2} + \frac{\beta_4}{2} }  } }   \frac{   \mathfrak q^{ \ff{-\pN/\nN}}   }{1 -  \mathfrak q^{-1} } ~ , \nn\\ 
     & =  \sum_{n_x , n_y \in \mathbb Z} \sum_{\ell \in \mathbb N} e^{ - 2 \im  \epsilon  k_0 \comm{ \omega \p{ - \ell   +  \ff{-\pN/\nN} } + \tau n_x - n_y + r \p{ - \tau \frac{\alpha_3}{2} + \frac{\alpha_4}{2} }  + q_G \p{ a_0 - \tau \frac{\beta_3}{2} + \frac{\beta_4}{2} }  } }    ~ . 
\end{align}
The latter becomes the following infinite product:
\begin{align}
    \widetilde Z_+ & = \prod_{n_x , n_y \in \mathbb Z} \prod_{\ell \in \mathbb N} \comm{ \amf  \p{ -  \ell   +  \ff{-\frac{\pN}{\nN} } } + \tau n_x - n_y + \frac{r}{2} \p{ - \tau  \alpha_3  +  \alpha_4 }  + q_G \p{ a_0 - \tau \frac{\beta_3}{2} + \frac{\beta_4}{2} }  }^{-1}     ~ , \nn\\
    %
    %
    & = \prod_{n_x , n_y \in \mathbb Z} \prod_{\ell \in \mathbb N} \acomm{ \amf \comm{  \ell  + \frac{1}{2}\p{- 1 + \mathfrak b + \mathfrak c} } + \tau n_x - n_y + q_i^\Phi \gamma_i  }^{-1}    ~ ,
\end{align}
where $\mathfrak b$ and $\mathfrak c$ are those introduced in \cite{Inglese:2023wky,Inglese:2023tyc} with $\st=-1$. Moreover,  in the expression $q_i^\Phi \gamma_i = r  \gamma_R + q_G \gamma_G$ appearing   in $\widetilde Z_+$, the fugacity   $\gamma_G$ is the one defined in  (\ref{eq: chemicalpotentialsspindlet2tt}), whereas   $\gamma_R$   is 
\begin{align}
	& \gamma_R = \amf \frac{\chi  }{4} +  \tau \frac{\alpha_3}{2} - \frac{\alpha_4}{2} ~ , &  \alpha_3, \alpha_4 \in \mathbb Z   ~ ,
\end{align}
in agreement with (\ref{eq: rsymconstspindlet2anyt}) for $\st=-1$. Regularizing the product we again find  an elliptic Gamma function:
\begin{align}\label{eq: ZblockNat}
           \widetilde Z_+ & =  \frac{e^{ 2 \pi \im  \widetilde \Psi_+ } }{ \p{ z   q^{ \frac{1}{2}\p{ \mathfrak b - 1} } ; q , p }_\infty  \p{ z^{-1} \, p \, q^{  \frac{1}{2}\p{1 - \mathfrak b} }  ; q^{-1} , p }_\infty} = e^{ 2 \pi \im  \widetilde \Psi_+ }   \Gamma_e\p{ z   q^{ \frac{1}{2}\p{ \mathfrak b - 1} } ; q , p }  ~ , \nn\\ 
         \widetilde \Psi_+ & =  - \frac{1}{48 \tau \amf  }\comm{ 1 - 2 q_i^\Phi \gamma_i + \tau - \amf \p{\mathfrak c + \mathfrak b} } \nn\\
          & \times \acomm{ 2 \tau - 4 q_i^\Phi \gamma_i\p{ 1 - q_i^\Phi \gamma_i + \tau  } - 2 \amf \p{\mathfrak c + \mathfrak b}\p{1 - 2 q_i^\Phi \gamma_i + \tau} + \amf^2\comm{ \p{\mathfrak c + \mathfrak b}^2 - 1 } } ~ . 
\end{align}
The object $\widetilde Z_+$   in (\ref{eq: ZblockNat}) is the building block encoding the north-pole contribution to  $Z_{\text{1-L}}$  in the case of anti-twisted $\spindle\times T^2$, whereas $ Z_+$   in (\ref{eq: ZblockNat}) is the building block encoding the north-pole contribution to  $Z_{\text{1-L}}$  in the case of anti-twisted $\spindle\times T^2$, whereas $  Z_+$   in (\ref{eq: ZblockN}) is the north-pole  block for topologically twisted $\spindle\times T^2$. We can write a single expression for both  $Z_+$ and $\widetilde Z_+ $  by making use of the parameter  $\st$:
\begin{align}\label{eq: ZblockNanytwist}
           & Z_+^{(\st)} =   e^{ 2 \pi \im \Psi^{(\st)} _+}   \Gamma_e\p{ z^{-\st}  q^{ \frac{\st}{2}\p{1 - \mathfrak b} } ; q , p }  ~ , & Z_+^{(+1)} = Z_+ ~ , \qquad  Z_+^{(-1)} = \widetilde Z_+  ~ .
\end{align} 
Multiplying (\ref{eq: ZblockNanytwist})  by the dual south-pole block (\ref{eq: ZblockS}) gives 
\begin{align} 
          Z^{(\st)}_{\text{1-L}} = Z_+^{(\st)} Z_-   = e^{ 2  \pi \im  \Psi^{(\st)}   } \frac{  \Gamma_e\p{ z^{-\st}  q^{ \frac{\st}{2}\p{1 - \mathfrak b} } ; q , p }  } { \Gamma_e\p{ z^{-1} q^{\frac{1}{2}\p{ 1 + \mathfrak b  }} ; q , p   }  }    ~ , 
\end{align}
for the 1-loop determinant of a $\mathcal N=1$ chiral multiplet on $\spindle \times T^2$ with any R-symmetry twist on $\spindle$. This justifies the conjecture (\ref{eq: 1loopcmspindlet2anytwist}) for a chiral multiplet in an arbitrary representation $\mathfrak R_G$ of  a gauge or flavour group $G$. As usual, the one-loop determinant of a $\mathcal N=1$ vector multiplet is obtained by setting $\mathfrak R_G = {\rm adj}_G$ and $r=2$.


 \subsection{Branched Lens Index}

 \paragraph{BPS locus.} 

  The gauge field  
\begin{align}
    a & =  a_\varphi\p{\vartheta} \dd\varphi  + \frac{h}{n}  \dd\psi + a_t\p{\vartheta} \dd t  ~ , \qquad h=0, \dots, \p{n-1}  ~ , \nn\\
    \mathfrak f_G & = \frac{1}{2\pi}\int_\spindle \dd a = \frac{\mathfrak m}{\nN \nS} ~ , \qquad \exp\p{\im \oint_{S^1_\psi}a}=e^{2 \pi \im h/n} ~ , \qquad \exp\p{\im \oint_{S^1_t}a} \in U(1) ~ ,
\end{align}
fulfils the BPS equations  $\lambda|_{\rm BPS} = \widetilde\lambda|_{\rm BPS}= \delta \lambda|_{\rm BPS} = \delta \widetilde \lambda|_{\rm BPS} = 0$ if 
\begin{align}
	& a_t\p{\vartheta} = - \im \beta \p{  \nN  b_1^{-1}  +  \nS  b_2^{-1}  } a_\varphi\p{\vartheta} + a_0  ~ , & a_0 \in \mathbb C ~ ,
\end{align}
implying that   $\Phi_G = \iota_K\mathcal A$ on the BPS locus reads 
\begin{align}
      \Phi_G = \iota_K \mathcal A|_{\rm BPS} = \frac{\knorm}{ L \beta} \comm{   a_0 - \im \beta h \p{ \tN b_1^{-1} + \tS b_2^{-1} } }  ~ ,
\end{align}
and that the profile of  $D|_{\rm BPS}$ is
\begin{align}
       D|_{\rm BPS} =  \frac{  \nN b_1^{-1} \cot\vartheta - \nS b_2^{-1} \tan\vartheta }{ L^2 \fs} a'_\varphi\p{\vartheta}   ~ .
\end{align}
Finally, for generic values of $\Phi_{R,G}$ the BPS equations governing the chiral multiplet, characterized by $\psi|_{\text{BPS}} = \widetilde\psi|_{\text{BPS}} = \delta \psi|_{\text{BPS}} = \delta \widetilde \psi|_{\text{BPS}} = 0$, are readily solved by the trivial configuration $\phi|_{\text{BPS}} = \widetilde \phi|_{\text{BPS}} = F|_{\text{BPS}} = \widetilde F|_{\text{BPS}} = 0$, mirroring the behavior observed on $\spindle \times T^2$.

 
 \paragraph{One-loop determinant.}

On $\blens\times S^1$, the eigenfunctions  $ \Phi \in {\rm ker}  L_{\widetilde Y}$ are regular in the neighbourhoods of   $\vartheta = 0$ and $\vartheta = \pi/2$ if  
\begin{align}
        q_G \mN - \nN n_\varphi + \tN \p{ n \, n_\psi - q_G h} \geq 0   ~ , \nn\\
           q_G \mS - \nS n_\varphi + \tS \p{ n \, n_\psi - q_G h}  \geq  0   ~ , 
\end{align}
which are inequalities satisfied by
\begin{align}
	n_\varphi & = \tS\p{ q_G \mN - n \ell_1 - j_1 } +  \tN\p{ - q_G \mS + n \ell_2 + j_2 } ~ ,  \nn\\
	 n_\psi & = \nN \ell_2 - \nS \ell_1 + \ff{ \frac{q_G\p{h - \mathfrak m} + \nN j_2 }{n} } - \ff{ \frac{\nS j_1 }{n} } ~ ,  \nn\\
	 \rem{\nS j_1}{n} & = \rem{q_G\p{h - \mathfrak m} + \nN j_2}{n} ~ , 
\end{align}
with $\ell_1, \ell_2 \in \mathbb N$ and $ j_1, j_2 = 0, \dots, \p{n-1}$.  Analogously, the eigenfunctions  $ B \in {\rm ker}  L_{ Y}$ are non-singular  near   $\vartheta = 0$ and $\vartheta = \pi/2$ if  
\begin{align}
        - q_G \mN + \nN m_\varphi + \tN \p{q_G h -  n \, n_\psi } \geq 0   ~ , \nn\\
           -q_G \mS + \nS m_\varphi + \tS \p{ q_G h - n \, m_\psi }  \geq  0   ~ , 
\end{align}
which are inequalities satisfied by
\begin{align}
	m_\varphi & = \tS\p{ q_G \mN + n \, k_1 + i_1 } -  \tN\p{ q_G \mS + n \, k_2 + i_2 } ~ ,  \nn\\
	 m_\psi & =   \nS k_1 - \nN k_2 + \ff{ \frac{q_G\p{h - \mathfrak m} + \nS i_1 }{n} } - \ff{ \frac{\nN i_2 }{n} } ~ ,  \nn\\
	 \rem{\nN i_2}{n} & = \rem{q_G\p{h - \mathfrak m} + \nS  i_1}{n} ~ , 
\end{align}
with $k_1, k_2 \in \mathbb N$ and $ i_1, i_2 = 0, \dots, \p{n-1}$. Altogether we have
\begin{align}
	Z_{\text{1-L}}^{\rm CM} = \prod_{j=0}^{n-1} \prod_{n_3 \in \mathbb Z} \prod_{\ell_1, \ell_2 \in \mathbb N} \frac{ \prod_{i \in\mathbb J_+\p{j, \mathfrak h } } \comm{ \amf_1\p{ \ell_1 + \frac{j}{n}  } + \amf_2\p{ \ell_2 + \frac{i}{n}  } + n_3 - q_G \gamma_G +\p{2-r} \gamma_R } }{ \prod_{k \in\mathbb J_-\p{j, \mathfrak h }} \comm{ \amf_1\p{ \ell_1 + \frac{k}{n}  } + \amf_2\p{ \ell_2 + \frac{j}{n}  } + n_3 + q_G \gamma_G + r \gamma_R } } ~ , 
\end{align}
with $\mathfrak h = \rem{h-\mathfrak m}{n}$ and
\begin{align}
	& \amf_{1,2}  = - \im n \beta b_{1,2}^{-1}    ~ , \nn\\
	& \gamma_R  = \frac{\amf_1 + \amf_2}{2 n} - \frac{\alpha_4}{2} ~ ,  \qquad \gamma_G  =  - \frac{\amf_1 \mN + \amf_2 \mS}{2 n} - a_0 ~ , \nn\\
	& \mathbb J_\pm\p{j, \mathfrak h} = \acomm{j_0=0, \dots, \p{n-1} : \rem{n_\pm j_0 }{n}  = \rem{q_G \mathfrak h + n_\mp  j}{n} }   ~ . 
\end{align}
We can regularize the infinite product above  by noticing that
\begin{align}
	 P_3\p{u ; \amf_1 , \amf_2} = \prod_{n_3 \in \mathbb Z}  \prod_{n_1 , n_2 \in \mathbb N}  \frac{1}{ \amf_1 n_1 + \amf_2 n_2 + n_3 + u  } ~ , 
\end{align}
can be expanded as
\begin{align}
	 P_3\p{u ; \amf_1 , \amf_2} & =  \prod_{n_1 , n_2, n_3 \in \mathbb N}  \frac{-1}{ \p{ \amf_1 n_1 + \amf_2 n_2 + n_3 + u  } \p{ - \amf_1 n_1 - \amf_2 n_2 + n_3 + 1 -  u  } } ~ ,  \nn\\
	 & =   \prod_{n_1 , n_2 , n_3 \in \mathbb N}  \frac{-1}{ \p{ - \amf_1 n_1 - \amf_2 n_2 + n_3 - u }  \p{ \amf_1 n_1 + \amf_2 n_2 + n_3 + 1 + u } } ~ ,
\end{align}
which can be written as a single object by introducing a sign $s=\pm1$
\begin{align}
	 P_3^{\p{s}}\p{u ; \amf_1 , \amf_2} & =  \prod_{n_1 , n_2, n_3 \in \mathbb N}  \frac{1}{ \p{ s \amf_1 n_1 + s \amf_2 n_2 + n_3 + s u  } \p{ - s \amf_1 n_1 - s \amf_2 n_2 + n_3 + 1 - s  u  } } ~ , \nn\\  
	  & =     \Gamma_3 \p{ s u | s \amf_1  ,  s \amf_2, 1 } \Gamma_3 \p{ 1 - s  u | - s \amf_1 , - s \amf_2  , 1   }  ~ ,   \nn\\
	 & =     \frac{e^{- \pi  \im  \zeta_3\p{0, s u , | s \amf_1 , s \amf_2} }}{  \p{ e^{2 \pi \im s u } ; e^{2 \pi \im s \amf_1 } , e^{2 \pi \im s \amf_2 }  } }    ~ ,  
\end{align}
where we used \cite{FELDER200044,FRIEDMAN2004362}. In summary,
\begin{align}\label{eq: reginfprodnnz}
	 \prod_{n_3 \in \mathbb Z}  \prod_{n_1 , n_2 \in \mathbb N}  \frac{1}{ \amf_1 n_1 + \amf_2 n_2 + n_3 + u  } \qquad \to \qquad   \frac{e^{- \pi  \im  \zeta_3\p{0, s u , | s \amf_1 , s \amf_2} }}{  \p{ e^{2 \pi \im s u } ; e^{2 \pi \im s \amf_1 } , e^{2 \pi \im s \amf_2 }  } }    ~ ,  
\end{align}
yielding  
\begin{align}
	Z_{\text{1-L}}^{\rm CM}  & = \prod_{\rho\in\mathfrak R_G} e^{2 \pi \im \Psi^{\rm CM} \comm{\rho\p{\gamma_G} , \rho\p{\mathfrak h}} } \prod_{j=0}^{n-1}  \frac{ \prod_{i \in \mathbb J_+\p{j, \rho\p{\mathfrak h} } } \p{ q_1^{  \p{1+j}/n  } q_2^{ \p{1+i}/n  }    z^{-\rho} ; q_1, q_2  }_\infty }{ \prod_{k \in \mathbb J_-\p{j, \rho\p{ \mathfrak h} }} \p{ q_1^{  k/n} q_2^{ j/n} z^\rho  ; q_1 , q_2 }_\infty } ~ , \nn\\
	\Psi^{\rm CM}\comm{\rho\p{\gamma_G} , \rho\p{ \mathfrak h}} & = \frac12  \sum_{j=0}^{n-1}  \sum_{i \in \mathbb J_+\p{j, \rho\p{\mathfrak h} } }  \zeta_3\p{ 0 ,    \frac{ \amf_1 j + \amf_2 i}{n}  -   \rho\p{\gamma_G} +\p{2-r} \gamma_R  | \amf_1 , \amf_2  } \nn\\
	& - \frac12  \sum_{j=0}^{n-1} \sum_{k \in \mathbb J_-\p{j, \rho\p{\mathfrak h} } }  \zeta_3\p{0,   \frac{ \amf_1 k + \amf_2 j}{n} +  \rho\p{\gamma_G} + r \gamma_R  | \amf_1 , \amf_2 }  	~ , 
\end{align}
where we defined the fugacities 
\begin{align}
		& q_{1,2} = e^{2 \pi \im \amf_{1,2} } ~ , & z = e^{2 \pi \im \p{ q_G \gamma_G + r \gamma_R} } ~ .
\end{align}
Correspondingly, the   vector-multiplet one-loop determinant is
\begin{align}
	Z_{\text{1-L}}^{\rm VM}  & = \prod_{\alpha\in {\rm adj}_G} e^{2 \pi \im \Psi^{\rm CM} \comm{\alpha\p{\gamma_G} , \alpha\p{\mathfrak h}} } \prod_{j=0}^{n-1} \left.  \frac{ \prod_{i \in \mathbb J_+\p{j, \alpha\p{\mathfrak h} } } \p{ q_1^{  \p{1+j}/n  } q_2^{ \p{1+i}/n  }    z^{-\alpha} ; q_1, q_2  }_\infty }{ \prod_{k \in \mathbb J_-\p{j, \alpha\p{ \mathfrak h} }} \p{ q_1^{  k/n} q_2^{ j/n} z^\alpha  ; q_1 , q_2 }_\infty } \right|_{r=2} ~ .
\end{align}


   


\bibliographystyle{JHEP}
\bibliography{4d_N1_susy}

\end{document}